\documentclass[aps,prb,preprint]{revtex4-2}
\usepackage{graphicx}
\usepackage{amsmath}
\usepackage{hyperref}
\usepackage[version=4]{mhchem}
\usepackage{gensymb}

\usepackage{siunitx}
\newcommand{\degC}{$^\circ$C}

\begin{document}


\title{Development of an Atomic Layer Deposition System for Deposition of Alumina as a Hydrogen Permeation Barrier} 



\author{Zachary R.\ Robinson}
\affiliation{Laboratory for Laser Energetics, University of Rochester. 250 E. River Rd, Rochester, NY 14623}
\email[]{zachary.robinson@rochester.edu}

\author{Jeffrey Woodward}
\affiliation{U.S. Naval Research Laboratory, Electronics Science and Technology Division. 4555 Overlook Ave. SW, Washington, DC 20375}

\author{James Michels}
\affiliation{SUNY Brockport, Department of Physics. 350 New Campus Drive, Brockport, NY 14420}

\author{Alexander C. Kozen}
\affiliation{University of Vermont, Department of Physics. 82 University Place, Burlington, VT 05405}

\author{Josh Ruby}
\author{Tyler Liao}
\author{Luke Herter}
\author{Rashad Ahmadov}
\author{Mark D. Wittman}
\author{Matt Sharpe}
\affiliation{Laboratory for Laser Energetics, University of Rochester. 250 E. River Rd, Rochester, NY 14623}


\date{\today}

\begin{abstract}

Tritium permeation into and through materials poses a critical challenge for the development of nuclear fusion reactors. Minimizing tritium permeation is essential for the safe and efficient use of available fuel supplies. In this work, we present the design, construction, and validation of custom atomic layer deposition (ALD) and deuterium permeation measurement systems aimed at developing thin-film hydrogen permeation barriers. Using the ALD system, we deposited conformal \ce{Al2O3} films on copper foil substrates and characterized their growth behavior, morphology, and composition. ALD growth rates of $\sim$1.1\AA/cycle were achieved for tempreatures between 100\degC{} and 210\degC. Permeation measurements on bare and ALD-coated copper foils revealed a significant reduction in deuterium flux with the addition of a $\sim$10~nm \ce{Al2O3} layer. While bare copper followed diffusion-limited transport consistent with Sievert’s law, the ALD-coated samples exhibited surface-limited, pore-mediated transport with linear pressure dependence. Arrhenius analysis showed distinct differences in activation energy for the two transport regimes, and permeation reduction factors (PRFs) exceeding an order of magnitude were observed. These results demonstrate the potential of ALD-grown \ce{Al2O3} films as effective hydrogen isotope barriers and provide a foundation for future studies on film optimization and integration into fusion-relevant components.

\end{abstract}


\maketitle 

\pagebreak

\section{Introduction}
Nuclear fusion holds immense promise for the future energy needs of our planet. In order to successfully build a nuclear fusion power plant, many material science challenges must be overcome \cite{knaster2016materials, strikwerda2024tritium}. In particular, we need new materials that can withstand the harsh environment of the reactor, contain the fusion fuel supply, and are straightforward to manufacturer at industrial scale. 

Currently, the amount of tritium available will be a significant limiting factor on the size and scale of nuclear fusion. Active work is ongoing to design fusion reactors such that they can breed their own tritium, utilizing the same fusion reaction that ultimately generates electricity. In the meantime, preserving as much existing tritium as possible is a high priority for the fusion energy industry. This challenge is complicated by the fact that tritium - as with all hydrogen isotopes - permeates most metals that would be used to contain gases in industrial systems, such as stainless steels. Tritium in particular can cause many problems once it permeates a metal, since it is hazardous, its decay product (\ce{^3He}) compromises the structural properties of the metal, and it can also result in hydrogen embrittlement for components at elevated temperature \cite{brass1994helium, robinson1991accelerated}. Additionally, any tritium atom that permeates into its container can't be easily retrieved and reinserted into the fueling system.

Thin, conformal barriers made of materials such as alumina (\ce{Al2O3}) have been extensively studied as hydrogen permeation barriers for many years \cite{serra2005hydrogen, levchuk2004deuterium, he2013influence}. The permeation reduction factor (PRF) is the factor by which the permeability of a material is reduced when a film is deposited on one surface. For \ce{Al2O3} films deposited on steel substrates, the PRF has been shown to have values up to 1000 for 1$\mu$m thick coatings on fusion-relevant Eurofer steel \cite{levchuk2004deuterium}. 

Atomic layer deposition (ALD) is a powerful technique for depositing thin, conformal films on substrates with arbitrary morphology \cite{george2010atomic}. ALD is a surface-mediated gas-phase deposition technique, and thus is not limited by line-of-sight the way that sputter deposition or thermal evaporation are. ALD is also a relatively low temperature technique that can be tuned to have excellent uniformity and thickness control \cite{george2010atomic, groner2004low}. Importantly, ALD could be used to deposit films on interior surfaces of various containers for nuclear fusion applications. Additionally, \ce{Al2O3} is both well-studied as a high quality material that can be deposited by ALD, and also has been shown to be an effective permeation barrier for hydrogen \cite{nemanivc2019hydrogen,serra2005hydrogen}.

In this work, we report the development of home-built ALD and permeation systems capable of depositing and characterizing the hydrogen permeation barrier performance of \ce{Al2O3} thin-films. This paper is part of a materials-focused program at the Laboratory for Laser Energetics exploring the usefulness of ALD to the fusion industry.

In the following sections, we present a detailed explanation of our home-built ALD and \ce{D2} permeation measurement systems. We have included many details of how we designed and built both systems. Our initial efforts were to grow ALD films on Si wafers, but quickly transitioned to copper substrates. Copper was chosen both for its chemical inertness with respect to hydrogen, along with its catalytic properties towards growth of graphene and other 2D materials \cite{binns2019structural, li2009large, feigelson2015growth}, which will be the topic of future work. Importantly, Cu is an excellent substrate because it has similar permeation behavior as some stainless steels, but in a single-element system that enables fundamental material science understanding \cite{steward1983review}.

\section{Experimental}
\subsection{ALD System}
In order to deposit \ce{Al2O3} on various substrates for evaluation as a hydrogen permeation barrier, a custom atomic layer deposition (ALD) system was developed. ALD was chosen because the films will ultimately need to be deposited on the interior surface of a canister used for tritium storage and transport. The eventual deposition in cylindrical cans was a primary design consideration, although for this preliminary study only planar substrates were used.

\begin{figure}[ht]
    \centering
    \frame{\includegraphics[width=0.7\linewidth]{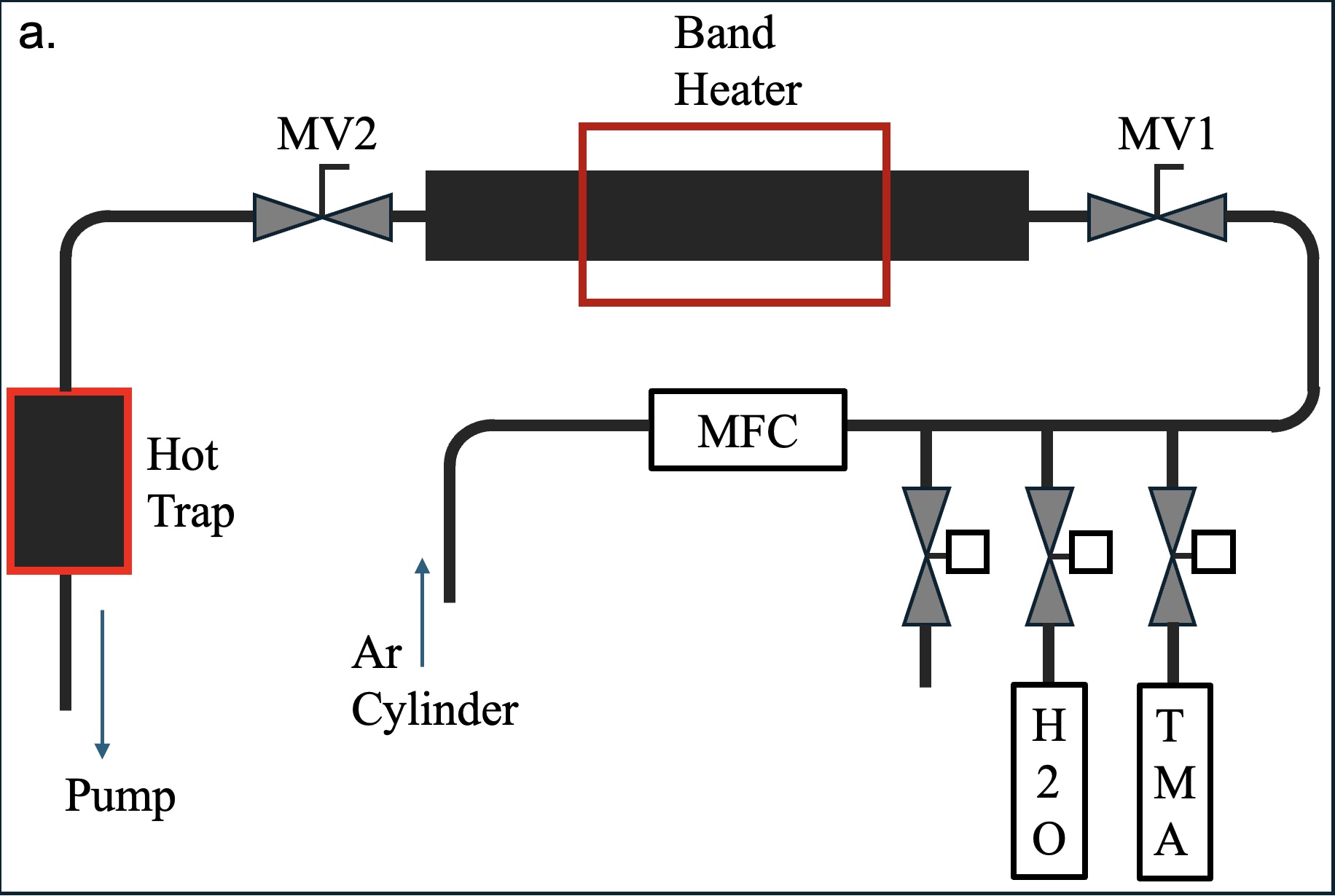}}
    \frame{\includegraphics[width=0.7\linewidth]{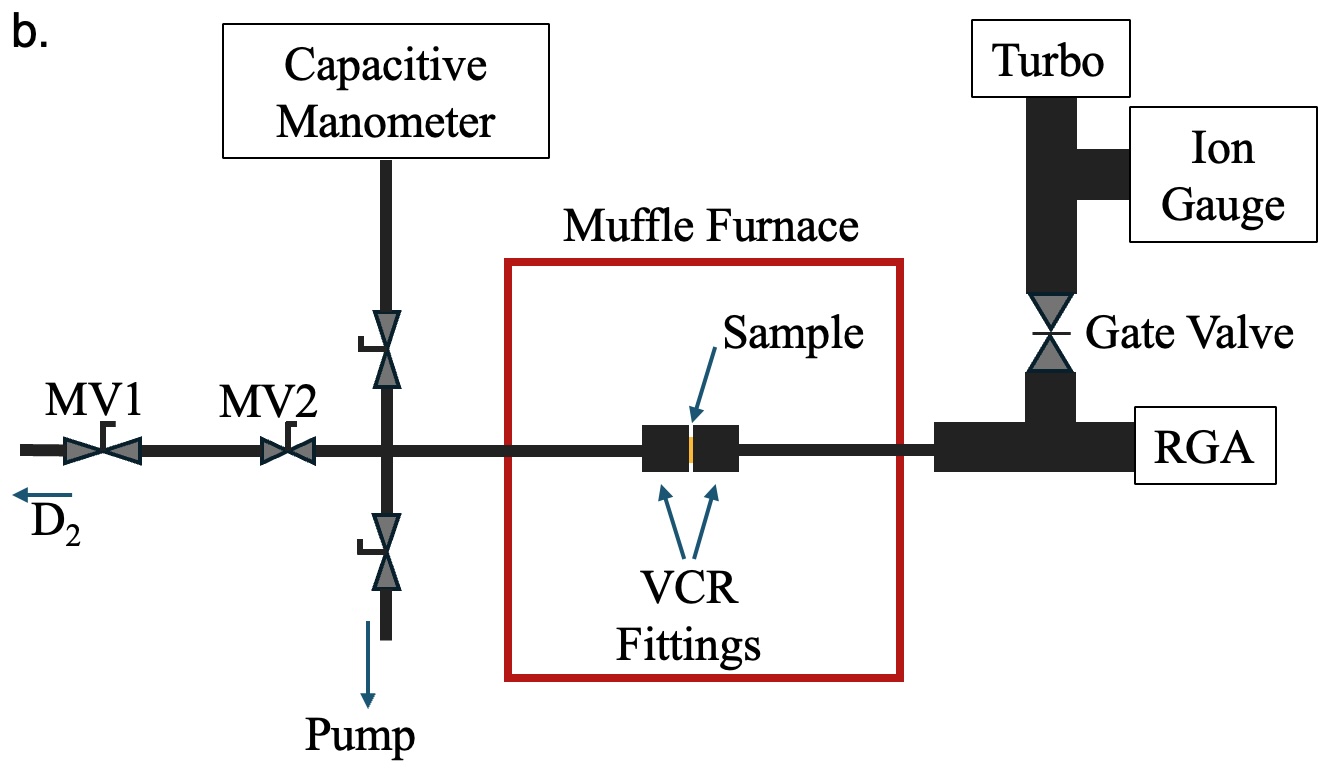}}
    \caption{Schematic representation of the atomic layer deposition system (a) and permeation system (b).}
    \label{fig:schematic}
\end{figure}

A schematic of the ALD system can be found in Figure \ref{fig:schematic}a. A gas cylinder containing an Ar carrier gas is connected to a mass flow controller (Alicat MC-50sccm), which allows for regulated and consistent gas-flow in our system. The Ar passes over a manifold with capacity for 3 ALD precursors, of which we had trimethyl aluminum (TMA), \ce{H2O}, and one unused port. The ALD valves are Swagelok ALD3, and are controlled by 24V MAC electro-pneumatic valves operating on house nitrogen supply at 60psi. 

The carrier gas and precursors flow through our chamber, which is 1" OD SS304 tube, connected to the reactor on each end with an ultra-Torr fitting adapted to a Swagelok 4H valve (see valves MV1 and MV2 in Figure \ref{fig:schematic}a). The distance between the 4H valves is approximately 30~cm. The chamber is heated with two 2" long band heaters, which provide an approximately 4" hot zone in the furnace. Finally, there is a hot trap (maintained at $\sim$200\degC) about 16" from the top of the rotary-vane pump.  

The primary electronics for the system are LabVIEW data acquisition cards (DAQs), which interface with all of the heaters, electro-pneumatic valves, and thermocouples. For the heaters, digital output channels in the DAQ are connected to solid state relays, which provide power to system heaters installed on all of the gas lines between the MFC and the pumping system. The 2 band heaters on the main reactor are connected in parallel and controlled by a separate DAQ/solid state relay. Thermocouples are installed throughout the exterior of the system.  

The pressure is measured with a 1~Torr capacitance manometer, which is controlled by an MKS PR4000B. The capacitance manometer is installed after the hot trap, to minimize interactions between the precursor molecules and the transducer. A custom Python program (available upon request) controls all of the hardware during growths. 

An interlock is installed on the system, requiring thermal fuses, an emergency stop button, and a LabVIEW-controlled main power switch are all satisfied for the precursor pneumatic valves to be powered. Evaluation of the system and data related to its operation are in the Results and Discussion section.

\subsection{Permeation System}
A schematic of the permeation system can be found in Figure \ref{fig:schematic}b. The system is typical of hydrogen permeation systems, with several important design considerations. The system is divided in half, with a `high pressure' and `high vacuum' side separated by the substrate being measured. 

The high pressure side of the permeation instrument consists of a dry roughing pump, 1000~Torr capacitance manometer, and a \ce{D2} supply. During a typical run the high pressure side is pumped out to the base pressure of the pump ($\sim$40~mTorr), with both valves on the \ce{D2} line open (MV1 and MV2) to evacuate the tubing up to the \ce{D2} regulator. Once the sample is at the desired temperature, MV2 is closed and the \ce{D2} gas line is pressurized. The pump valve is then closed, and the gas between MV1 and MV2 is allowed to expand into the high pressure side. This procedure typically brings the high pressure side to approximately 80~Torr. The process of expanding the gas between MV1 and MV2 is then repeated for 5 or 6 different upstream pressures, allowing time for the downstream pressure to equilibrate in between pressure changes.

The high vacuum side of the permeation instrument consists of a residual gas analyzer (RGA) and a turbo pump with a mechanical roughing pump backing it. The RGA runs continuously during an experimental run, measuring the partial pressure of \ce{D2} as a function of time. The base pressure of the high vacuum side of the permeation system is in the low 10$^{-9}$~Torr range.

The sample itself resides inside a muffle furnace that is continuously purged with dry nitrogen during an experimental run. The samples, which for this study primarily consisted of 25~$\mu$m Cu foils with and without 10~nm ALD \ce{Al2O3}, are compressed between two copper VCR gaskets that are tightened in a 1/4" fitting. 

\subsection{Sample Preparation and Characterization}
All of our permeation measurement samples consisted of a copper foil substrate (Thermo Scientific Chemicals 0.001" Cu Foil, 99.8\%). The foil was measured as-received, annealed in our tube furnace, and annealed with an ALD film grown on top. The substrates were annealed in a vertically-oriented Mellen MTSC tube furnace that was continuously purged with 5\%~\ce{H2}, 95\%~\ce{Ar}. The gas was passed through Restek indicating oxygen and moisture filters prior to being diverted into the furnace. The gas was regulated to 50~sccm. All anneals were performed at 900\degC{} with a 90~minute heatup, 20~minute anneal, and several hours cool-down. The samples were approximately 0.5~inches in diameter. Photos of the as-received and annealed Cu foils can be found in the Supplementary Data. 

All films for ALD process development were grown on $\sim$1~cm~x~1~cm silicon wafer pieces. They were characterized with a JA Woollam RC2 ellipsometer. Prior to each scan the sample tilt and height were optimized using automated functions. Scans from 300~nm - 1000~nm were taken at angles of 65\degree, 70\degree, and 75\degree. The spectra were fit to a Cauchy \ce{Al2O3} with A~=~1.751, B~=~0.00632, and C~=~-0.00010152. The native oxide beneath the \ce{Al2O3} was characterized on a control substrate, and fixed for subsequent measurements. Characteristic data can be found in the Supplemental Data section.

A subset of the samples were measured with X-ray photoelectron spectroscopy using a Kratos Axis Ultra DLD. Monochromatic Al-k$\alpha$ X-rays were used. Survey scans were performed with a 100~eV pass energy, and high resolution scans were performed with a 20~eV pass energy. 

X-ray diffraction was also performed, primarily to characterize the crystallization of the copper foil substrates. Samples were measured with a Rigaku SmartLab XRD system, and analyzed with a combination of fitting and Fourier analysis. All fits were performed with the Rigaku GlobalFit software. The same Rigaku system was also used for X-ray Reflectivity, in order to measure the mass density of the films, and also as a comparison with the ellipsometry-based thickness measurements.

Atomic force microscopy (AFM) was performed to measure the film morphology. A Park Systems XE-70 microscope was used in tapping mode, and the data were analyzed using the Gwyddion software package.

\section{Results and Discussion}
\subsection{ALD}
Initial experiments depositing \ce{Al2O3} on silicon substrates yielded growth rates over 1.5\AA/cycle, which indicates a growth rate that is outside of the typical ALD process window \cite{ham2022investigation, OTT1997135}. This growth rate persisted down to 20~ms pulses, which is the shortest pulse we could reliably measure from our ALD valves (pulse-timing was measured with the voice recorder app on an iPhone, as shown in Supplementary Data), and approaches the limit of the valve operation specification. Therefore, blank VCR gaskets with nominally 250~$\mu$m holes were installed between the ALD precursor manual-valves and the manifold pneumatic valves. See the Supplemental Data for an optical microscope image of the ALD aperture. Subsequently, the saturation-curve growth study shown in Table \ref{tab:ALDGrowth} was performed. Note that we performed process development on as-received Si wafers with a native oxide for development of the system and comparison with literature. Growth on copper substrates for subsequent permeation measurements will be discussed later.

For all of the growths in Table \ref{tab:ALDGrowth}, the purge time was fixed at 30~s, the flow-rate of gas was fixed at 10~sccm, and the reactor temperature was 160\degC. Growth-per-cycle (GPC) of around 1.1\AA, which is typical of much of the ALD literature, was achieved for almost all of the precursor pulse times. This indicates that our reactor is operating well within the ALD process window.

\begin{table}
\begin{tabular}{|c|c|c|}
    \hline
     TMA (ms) & H$_2$O pulse (ms) & GPC ($\textnormal{\AA}$) \\\hline
     100 & 100 & 1.5 \\\hline
     20 & 100 & 1.2 \\\hline
     100 & 20 & 1.1 \\\hline
     50 & 50 & 1.2 \\\hline
     20 & 50 & 1.2 \\\hline
     50 & 20 & 1.1 \\\hline
     20 & 20 & 1.1 \\\hline
\end{tabular}
    \caption{ALD process window for 160\degC{} growth and 100 cycles, with a 30~s purge between each precursor pulse. All growths have a 250$\mu$m aperture installed between the precursor cylinder and the ALD valve to limit conductance into the growth reactor. Pulse timing is the nominal pulse.}
    \label{tab:ALDGrowth}
\end{table}

\begin{figure}
    \centering
    \frame{\includegraphics[width=0.7\linewidth]{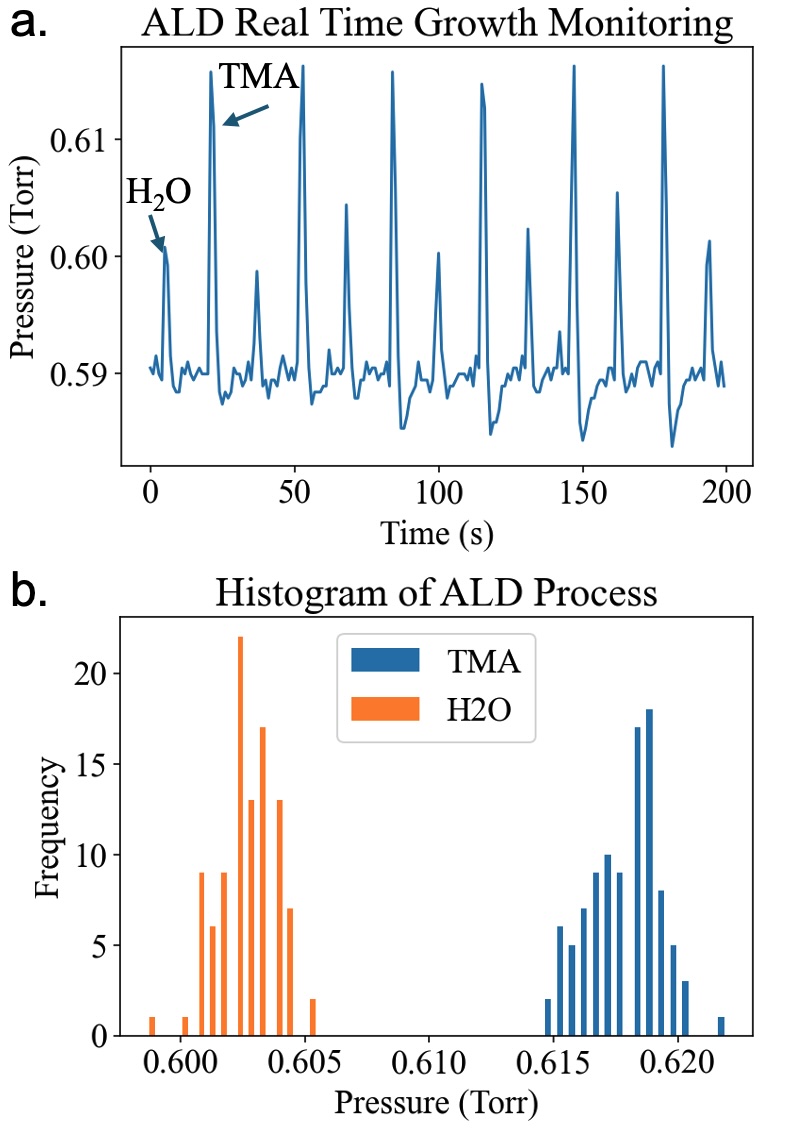}}
    \caption{Characterization of the pressure in the ALD system as a function of time during a typical growth process. The pressure was measured with a 1~Torr capacitance manometer to resolve peaks around 0.02~Torr tall. The top plot (a) is a plot of the raw log file. The bottom plot (b) is a histogram generated from the peak intensity of each peak in plot a. Peaks were identified using the findpeaks function in SciPy python library.}
    \label{fig:ALDsystem}
\end{figure}

In order to characterize our ALD growth reactor, the Python-based control program records the pressure approximately every 500~ms (See Figure \ref{fig:ALDsystem}a). Each pulse was then identified using a peak-identifying library in Python, and a histogram of the \ce{H2O} and TMA peak-pressure was generated (see Figure \ref{fig:ALDsystem}b). The histogram provides a diagnostic following each growth, and changes can be monitored over time as the system is operated. 

\begin{figure}
    \centering
    \includegraphics[width=0.7\linewidth]{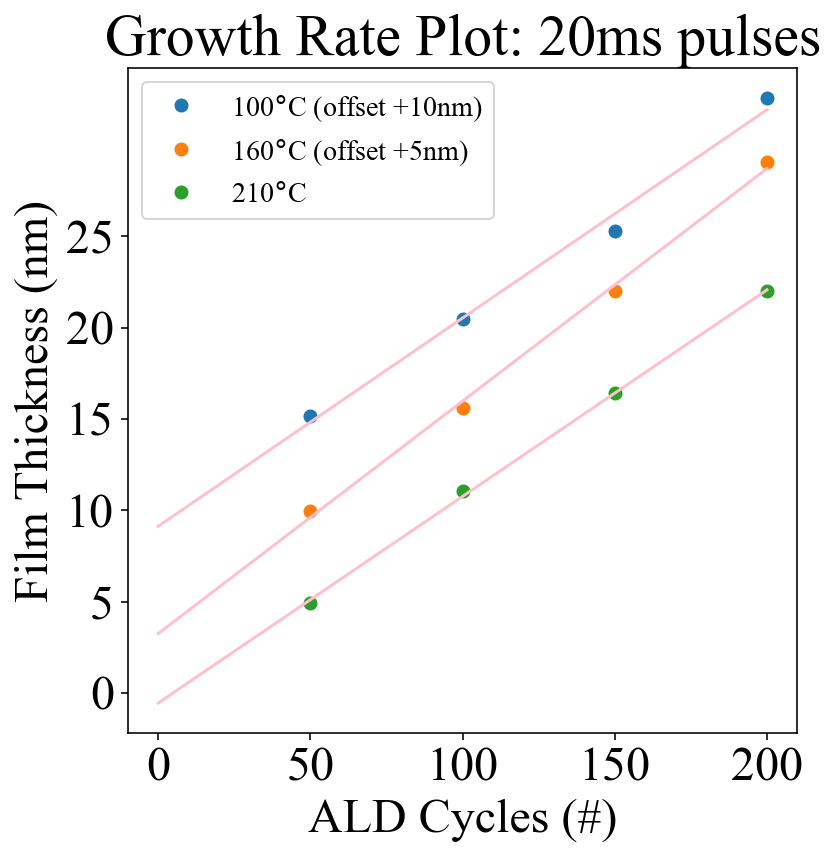}
    \caption{Plot of ALD film thickness as measured with ellipsometry as a function of the number of ALD cycles for three different growth temperatures. Note that the data are offset for clarity: 160\degC{} +~5nm, 210\degC{} +~10nm. The growth rate (slope) and incubation period are shown in Table \ref{tab:gpc}. These growths were performed on Si wafers with a native oxide thickness of about 3nm.}
    \label{fig:ALDgpc}
\end{figure}

Once the growth process was established to be within the typical ALD process window, the growth rate was measured at four temperatures by performing 50, 100, 150, and 200 cycle growths. A linear fit was added to each set of fixed-temperature growths, as shown in Figure \ref{fig:ALDgpc}. Note that in the figure the data are offset for clarity. 
A least-squares fit was performed using Excel's LINEST function, with the results shown in Table \ref{tab:gpc}. Both the GPC and incubation period (y-intercept) were extracted from the fit, along with their corresponding uncertainty. Note that the spatial distribution of thickness across our reactor was not measured in this work because the `hot zone' in our ALD reactor is prohibitively small. 

\begin{table}
    \centering
    \begin{tabular}{|c|c|c|}\hline
    
        Growth Temp (\degC{}) & GPC (\AA/cycle) & y-intercept (\AA) \\\hline
       100  & 1.14 $\pm$ 0.07 & -9 $\pm$ 10 \\\hline
       160  & 1.27 $\pm$ 0.05 & -18 $\pm$ 6 \\\hline
       210  & 1.13 $\pm$ 0.02 & -5.6 $\pm$ 0.3 \\\hline
    \end{tabular}
    \caption{Growth per cycle (GPC) extracted from the slope of the film thickness vs cycle number plot shown in Figure \ref{fig:ALDgpc}. The uncertainty column results from a first-order least-squares fit of the data.}
    \label{tab:gpc}
\end{table}

\begin{figure}
    \centering
    \includegraphics[width=0.5\linewidth]{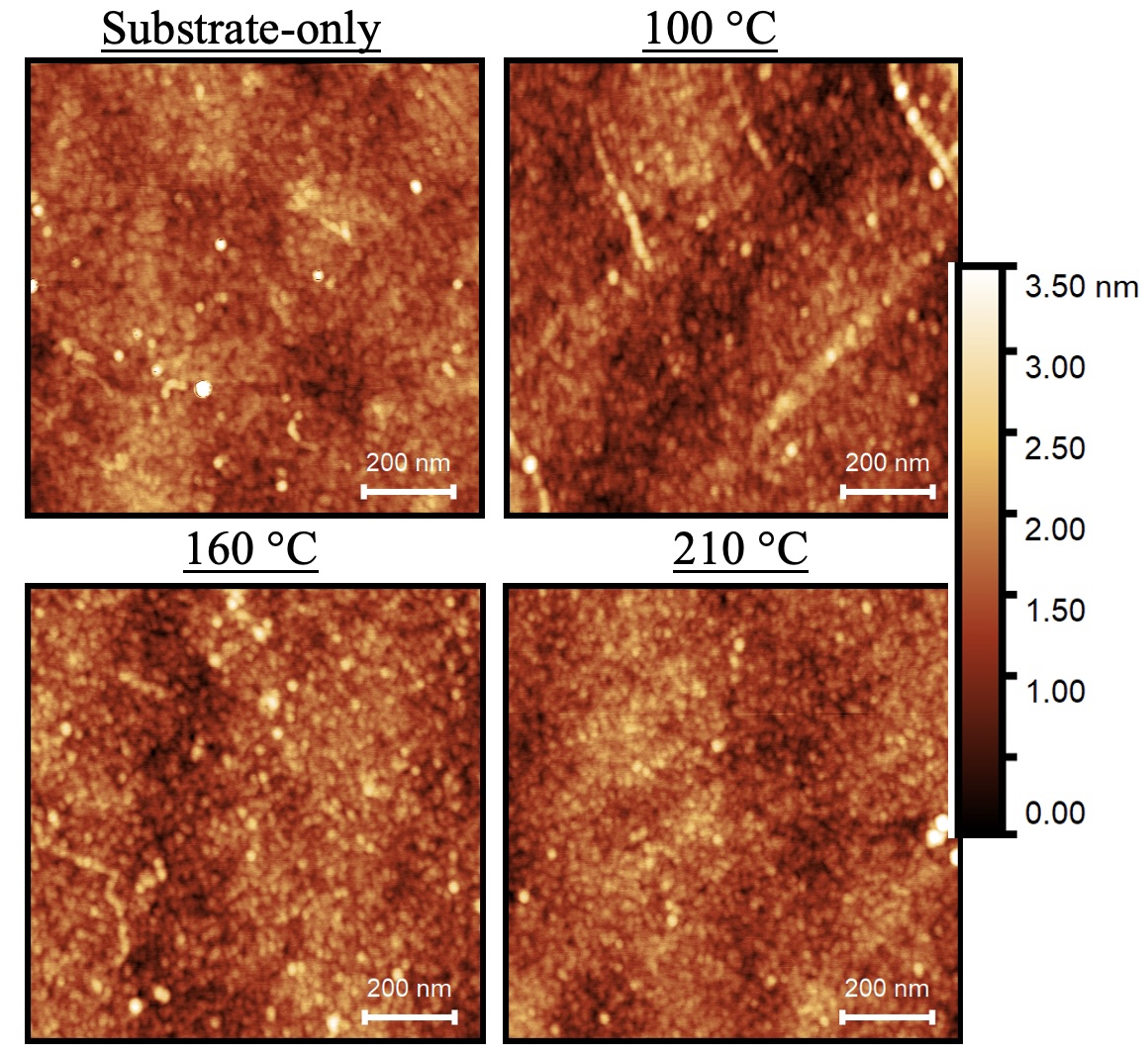}
    \caption{AFM images of the a) as-received Si substrate, b) 100\degC{} film, c) 160\degC{} film, and d) 210\degC{} film. The surface roughness values are given in the text.}
    \label{fig:AFM}
\end{figure}

The AFM images for the 200-cycle ALD films grown on Si can be found in Figure \ref{fig:AFM}. All of the images are qualitatively similar. The images were analyzed with Gwyddion following a plane-fit, and statistical parameters were extracted for the full 1$\mu$m~x~1$\mu$m images. RMS roughness values of 3.4\AA, 4.1\AA, 4.0\AA, and 3.6\AA, were obtained for the control sample and the samples grown at 100\degC, 160\degC, and 210\degC, respectively.

\begin{figure}
    \centering
    \includegraphics[width=0.5\linewidth]{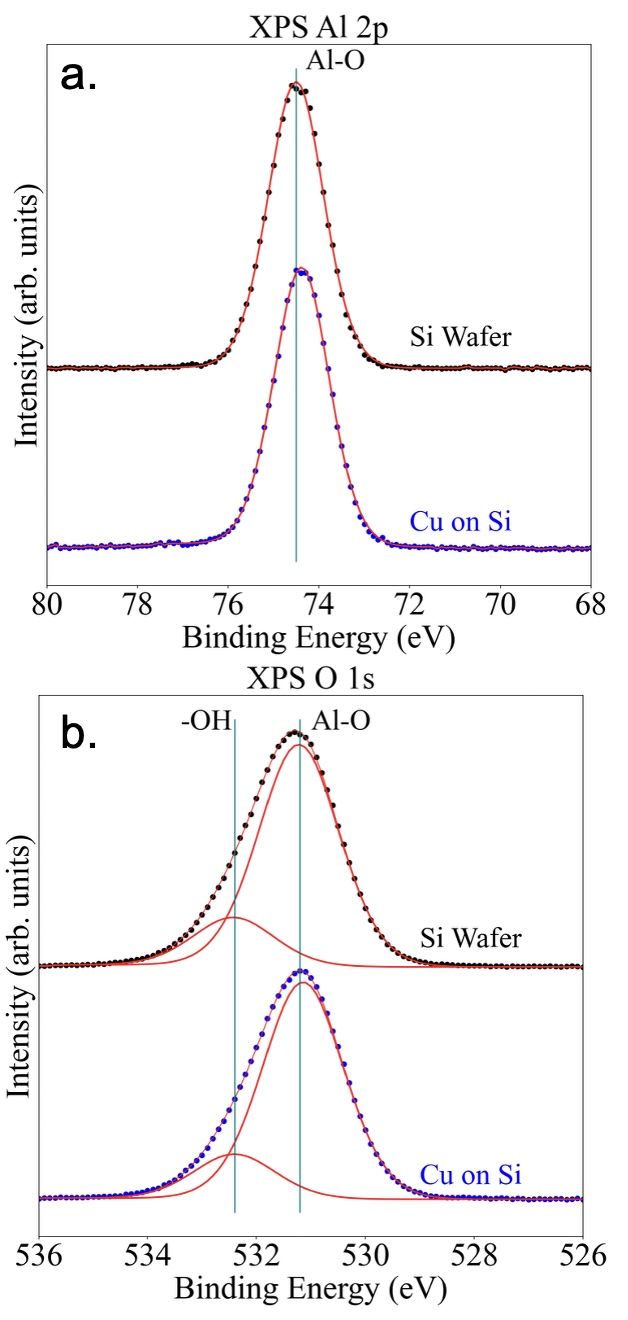}
    \caption{XPS data showing the Al 2p and O 1s binding energy ranges for an ALD deposition on both the Si substrate and on a 100~nm Cu film on a Si wafer. Spectra were measured for 100-cycle growths at all temperatures in this study. XPS spectra show typical \ce{Al2O3} behavior, with no impurities observed in survey spectra.}
    \label{fig:XPS}
\end{figure}

The XPS data can be found in Figure \ref{fig:XPS}. Spectra were measured for 100 cycle growths for all three temperatures, along with a control sample. An additional sample with a 100~nm e-beam evaporated Cu film was also measured following ALD growth at 160\degC, for comparison with films grown on Si. All of the samples were nearly identical, with only Al, O, and a small amount of adventitious carbon present in the survey scan. The measurements and peak fitting for the Al 2p binding energy and O 1s binding energy are shown in Figures \ref{fig:XPS}a and b, respectively, for both the Si wafer and the Si wafer with Cu. The O 1s:Al 2p ratios were calculated to be 1.4, 1.4, 1.5, and 1.4 for growths at 100\degC{} (Si), 160\degC{} (Si), 210\degC{} (Si), and 160\degC{} (Cu on Si), respectively. 

\begin{figure}
    \centering
    \includegraphics[width=0.5\linewidth]{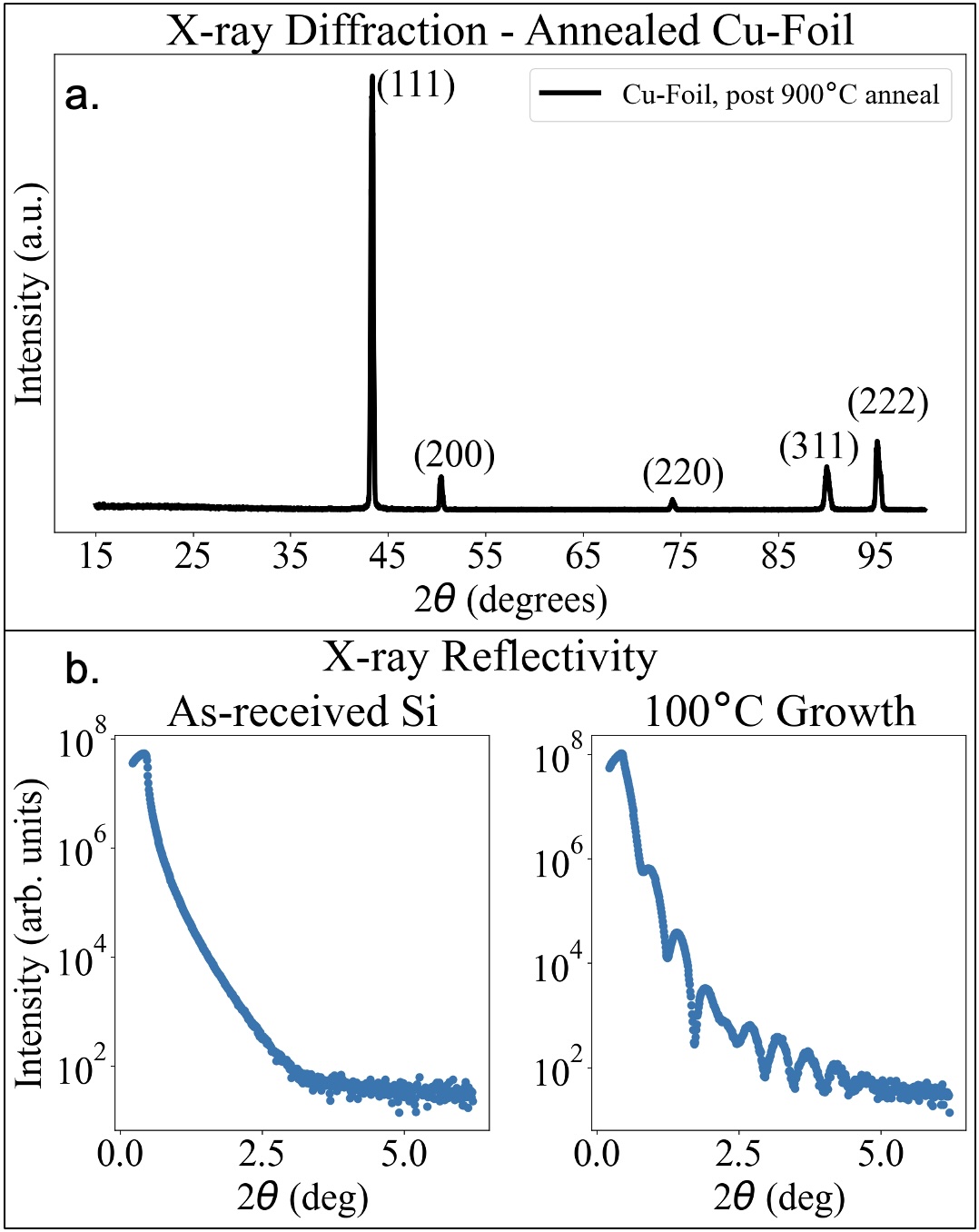}
    \caption{X-ray Diffraction (a) showing a crystalline Cu-foil following the 900\degC{} anneal. Note that the peak intensity ratios imply that the Cu-foil has a preferred orientation, likely in the (100) surface termination. X-ray reflectivity shown in (b) for the substrate-only (left) 100\degC{} ALD growth (right).}
    \label{fig:XRR}
\end{figure}

A subset of samples were also measured using XRD and XRR. The XRD measurement was performed on the annealed Cu-foil, to measure the crystallization state of the Cu following the 900\degC{} anneal. The data can be found in Figure \ref{fig:XRR}a. All of the peaks correspond to the FCC Cu lattice. The (111) and (222) peaks are shifted slightly, corresponding to a $\sim$~1\% out-of-plane compressive strain. The strain in the Cu likely originates from the permeation measurement, where the sample was heated to 350\degC{} while clamped between two gaskets in a VCR fitting. The peak intensity ratio indicates a preferential orientation, as has been observed in other studies for annealed Cu foils \cite{robinson2012substrate,sharma2017influence}.

XRR was used primarily to confirm the thickness measurements made with ellipsometry. A control sample (Figure \ref{fig:XRR}b) and samples grown on Si with 150 cycles (representative plot shown in Figure \ref{fig:XRR}c) were measured. The thickness measurement comparison can be found in Table \ref{tab:thickness}, showing agreement to within 1~nm. We also report the XRR-measured mass density of the films, which agrees reasonably well with literature values of around 3~g/cm$^3$ \cite{groner2004low, van2007plasma}.

\begin{table}
 \begin{tabular}{|c|c|c|c|}
     \hline
      Sample & Ellipsometry: \ce{Al2O3} Thickness (nm) & XRR: \ce{Al2O3} Thickness (nm) & Mass Density (g/cm$^3$)\\\hline
      100\degC{} & 15.3 &  16.1 & 3.18\\\hline
      160\degC{} & 17.0 & 17.4 & 3.19\\\hline
      210\degC{} & 16.4 & 16.9 & 3.19\\\hline
 \end{tabular}
     \caption{Film thickness comparison between ellipsometry and X-ray reflectivity for 150 cycle ALD growth.}
     \label{tab:thickness}
 \end{table}

\subsection{Permeation}
All of the samples measured in our permeation system were on a 25~$\mu$m Cu foil. Copper was chosen because of its promise as a catalyst material for two-dimensional material studies, and because it doesn't readily form a hydride when exposed to hydrogen species. Therefore, it is an ideal single-component FCC metal for studying effects of changing various material properties such as the crystalline orientation of the surface, grain size, and the presence of barrier layers. It also has similar permeability to many stainless steels \cite{steward1983review}. 

Prior to measuring permeation of \ce{D2} through ALD-coated Cu, the substrates themselves were characterized. Both as-received and annealed copper foils were measured. When measuring hydrogen permeation, there are several possible modes of transverse transport from the high pressure side to the high vacuum side of the permeation instrument. For metals, the permeation is generally governed by Sievert's Law for bulk diffusion limited transport, 

\begin{equation}
    S = \frac{C}{P^{1/2}}.\\
    \label{eqn:sievert}
\end{equation}

In the above equation, C is the concentration of hydrogen (or deuterium, as is the case for our experiments) in the substrate, S is the solubility of \ce{D2} in the copper, and P is the \ce{D2} pressure on the Cu surface. This model presumes dissociation of molecular species on the upstream surface of the metal, followed by atomic diffusion through the metallic substrate. 

In some cases, surface effects (rather than bulk diffusion) limit the permeation, which results in linear pressure dependence rather than the square-root dependence in Equation \ref{eqn:sievert}. This will be discussed in more detail following the discussion of Sievert's Law.

Combining Sievert's Law with Fick's First Law in 1-D,

\begin{equation}
    J = -D\frac{dC}{dx}
\end{equation}

yields 

\begin{equation}
    J = -\Phi\frac{dP_{\textrm{D}_2}^{1/2}}{dx}.
\end{equation}

In this equation, J is the flux in units of mol/m$^2$s, and $\Phi$ is defined as the permeability (diffusivity~$\times$~solubility). The transverse permeation flux is across the membrane itself, so the derivative can be interpreted as the finite difference across the thickness of the foil,

\begin{equation}
    J = -\Phi\frac{\Delta P_{\textrm{D}_2}^{1/2}}{\Delta \textrm{x}}.
\end{equation}

The total molar flow rate across a foil with area A is given by

\begin{equation}
    JA = \frac{dQ}{dt} = -\Phi \frac{A}{\Delta \textrm{x}}((P_{\textrm{D}_2}^{1/2})_{\textrm{downstream}}-(P_{\textrm{D}_2}^{1/2})_{\textrm{upstream}}),
\end{equation}

where $\frac{dQ}{dt}$ is the molar flow rate (mol/s). In the above equation, the upstream pressure (${P_{D_2,}^{1/2}}_{\textrm{upstream}}$) is generally measured with a pressure gauge, the area and thickness are geometrical factors of the membrane itself. The downstream pressure, $(P_{D_2}^{1/2})_{\textrm{downstream}}$, is usually many orders of magnitude lower than the upstream pressure, and can be disregarded,

\begin{equation}
    JA = \frac{dQ}{dt} = \Phi \frac{A}{\Delta \textrm{x}}(P_{\textrm{D}_2}^{1/2})_{\textrm{upstream}}.
    \label{Fick}
\end{equation}

It should be noted that ultimately our goal is to measure the permeability, $\Phi$. Since the downstream RGA can only measure the partial pressure of the gas, and not the molar flow rate as needed, we need to convert the partial pressure of the permeating gas into $\frac{dQ}{dt}$, which has units mol/s. To do this, we take the derivative of the ideal gas law with respect to time, for stable downstream temperature and pressure (ie - once equilibrium has been reached),

\begin{equation}
    P\dot{V} = \dot{n}RT
\end{equation}

\begin{figure}
    \centering
    \includegraphics[width=0.5\linewidth]{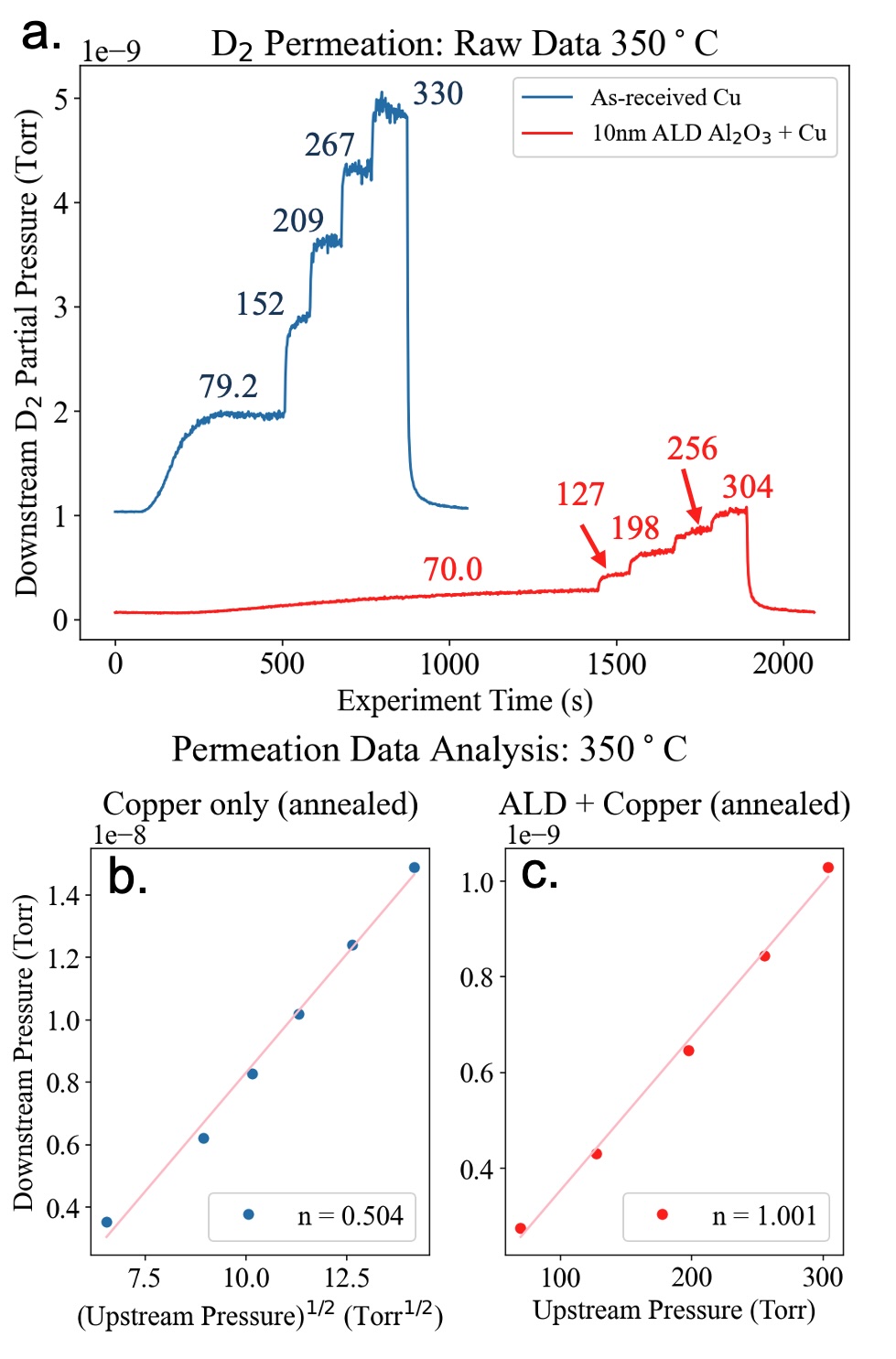}
    \caption{Plot a. is the raw data collected for the 350\degC{} permeation measurements on both as-received Cu and the ALD coated annealed Cu (data offset for clarity). The different plateaus in each plot represent a different upstream \ce{D2} pressure, which are indicated in units of Torr. Similar upstream pressures were used on both samples, despite the decreased downstream pressure apparent in the ALD data. Note that the ALD film took significantly longer to reach equilibrium with the first \ce{D2} exposure. The as-received data are offset in the y-direction for clarity. The bottom plots demonstrate the analysis, both for an annealed Cu sample (b.) and the ALD-coated Cu (c.) sample. Note the different x-axes on plots b. and c. The legend indicates the result of a fitting routine used to determine the exponent.}
    \label{fig:permdata}
\end{figure}

This equation assumes that the pressure and temperature have achieved steady-state on the downstream side, and that gas molecules permeating the system ($\dot{n}$) are pumped away with pumping speed $\dot{V}$. In that case, 

\begin{equation}
        \dot{n} = \frac{dQ}{dt} = \frac{P_{\textrm{downstream}}\dot{V}}{RT}
    \label{ideal}
\end{equation}

where T is the temperature of the gas being pumped away (ie - not the membrane temperature), and P$_\textrm{downstream}$ in this equation is the equilibrium pressure on the downstream side of the foil. Note that reaching equilibrium generally takes anywhere from a few seconds to much longer (up to many hours) depending on the thickness, area, temperature, and permeability of the membrane. Representative temporal data taken on an annealed Cu-foil substrate and an ALD-coated Cu-foil substrate can be found in Figure \ref{fig:permdata}a.

Substitution of equation \ref{ideal} into equation \ref{Fick} yields

\begin{equation}
    P_{\textrm{D2, downstream}} = \frac{\Phi A}{\dot{V} \Delta \textrm{x}}RT P{_{\textrm{D2,upstream}}^{1/2}}
\end{equation}

This equation was used to calculate the permeability of the metal-foil samples that obeyed Sievert's Law (i.e., showed downstream pressures that went as the square root of the upstream pressure).

Two copper-only substrates were measured in our permeation system prior to measuring a sample with an ALD \ce{Al2O3} film. The two samples measured were as-received, and one sample that was annealed in our tube furnace at 900\degC{} (see the Experimental section). The permeability of both samples was measured at temperatures of 350\degC, 325\degC, 300\degC, and 275\degC. Characteristic raw data are shown in Figure \ref{fig:permdata}. A linear fit was used to extract the permeability, as described above.

In order to verify the square-root dependence of the downstream pressure on upstream pressure, the data were fit with the equation $P_{\textrm{down}} = mP_{\textrm{up}}^n+b$, where m is the slope, n is the exponent, and b is the intercept. In this fit, $n$ describes the behavior that limits the permeation process, and informs our analysis. Data measured for the annealed Cu, and the annealed Cu with an ALD film at 350\degC{} are shown in Figure \ref{fig:permdata} b and c. The best-fit $n$ value is in the legend of each plot.

Fundamentally different limiting behavior was observed for the sample with ALD \ce{Al2O3} deposited on top of it. Figure \ref{fig:permdata}c shows that the downstream pressure is linearly dependent on the upstream pressure. We followed the analysis detailed in Ref \cite{young2021graphene}, since the behavior we observed was similar to the graphene samples measured in that study. Specifically, permeation through our sample was found to increase with increasing temperature, implying a pore-permeation model as opposed to a steric model \cite{yuan2019analytical, young2021graphene}. Therefore, Equation \ref{Fick} was modified such that

\begin{equation}
    \frac{dQ}{dt} = \frac{R_{ALD}A}{\Delta x}(P_{\textrm{D}_2})_{\textrm{upstream}}
    \label{eqn:linPerm}
\end{equation}

\begin{figure}
    \centering
    \includegraphics[width=0.7\linewidth]{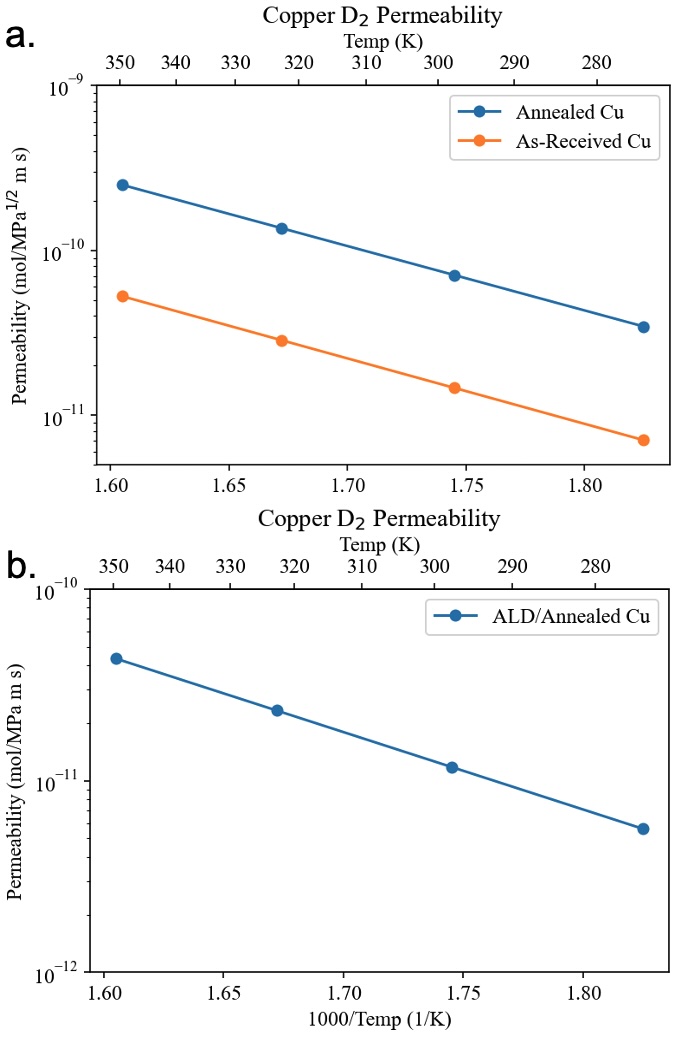}
    \caption{Arrhenius plot of Cu-foil samples (a) and ALD-coated Cu (b). Note the different y-axis units to account for the bulk vs surface limited limiting step.}
    \label{fig:Arrhenius}
\end{figure}

In this case, R$_\textrm{ALD}$ replaces the permeability, and has units of mol/(m~s~MPa). The temperature dependence of the permeability for the bare Cu samples and R$_\textrm{ALD}$ was extracted with an Arrhenius plot, as shown in Figure \ref{fig:Arrhenius}b. Due to the different permeation units, the bare Cu samples are plotted separately from the ALD-coated sample. In both cases, the prefactor and effective activation energies are extracted, and shown in Table \ref{tab:Permeation}. In the case of the bare copper, the prefactor and activation energy are related to the bulk diffusion through the copper. For the ALD-coated sample, the prefactor and activation energy are related to the defect density of the ALD film, and permeation through pores in the film \cite{young2021graphene}. For the bare Cu, both the prefactor and activation energy are similar to other values reported in the literature \cite{magnusson2017diffusion, steward1983review}.

\begin{figure}
    \centering
    \includegraphics[width=0.7\linewidth]{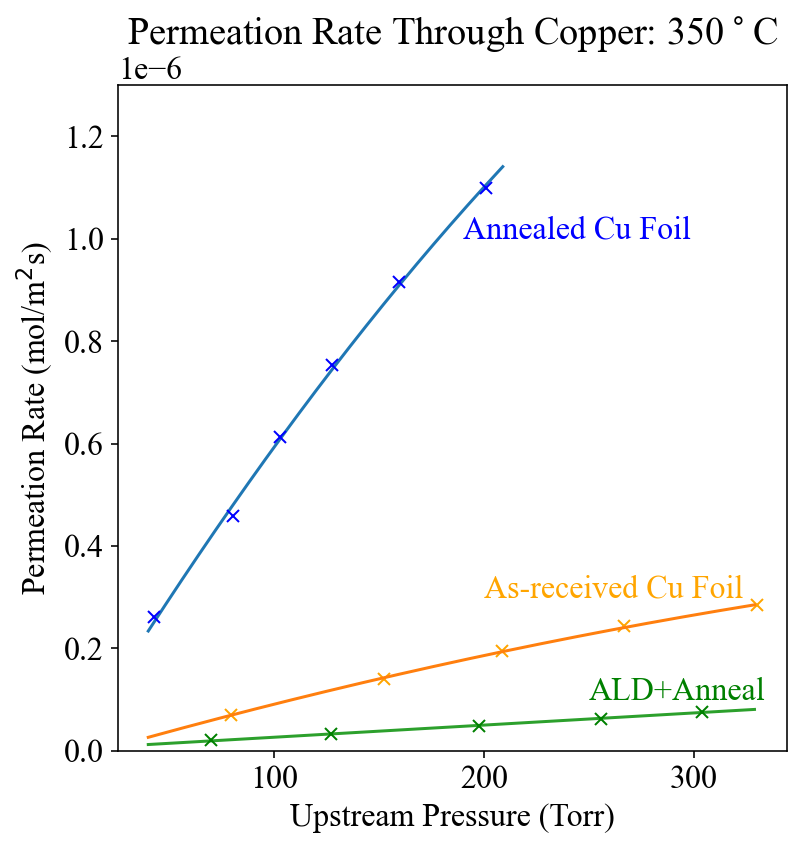}
    \caption{Comparison of permeation rates for as-received Cu, annealed Cu, and ALD-coated annealed Cu. Measurements performed at 350\degC.}
    \label{fig:permrate}
\end{figure}

In order to compare the ALD-coated Cu with the bare Cu substrates, the permeation rate (in units of mol/(m$^2$s), which is the steady-state permeation flux) was plotted for all three samples. These data are shown in Figure \ref{fig:permrate}. It can be seen that the addition of a $\sim$10~nm ALD \ce{Al2O3} film considerably impedes the rate of permeation, in addition to fundamentally changing the mechanism from bulk diffusion-limited permeation to surface limited permeation. It is important to note that the annealed Cu foil shows considerably higher permeation than the as-received copper. Presumably, this is because the annealed copper has a considerably thinner native oxide, has recrystallized with a faceted surface topography, and has more well-defined grain boundaries compared to the more amorphous as-received foil \cite{LEE2003102,robinson2012substrate}.

Typically, the permeation reduction factor is defined as

\begin{equation}
    PRF = \frac{\Phi_{\textrm{substrate}}}{\Phi_{\textrm{overlayer+substrate}}}.
\end{equation}

However, in our case, addition of the overlayer changes the mechanism limiting permeation (see Figure \ref{fig:permdata}b and c). Therefore, we calculate the permeation reduction factor with the permeation rate to remove the upstream pressure dependence from the analysis. All PRF values, which are shown as a function of temperature in Table \ref{tab:PRF}, were calculated for an upstream pressure of 100 Torr.  It is important to note that the PRF values represent the effect of a $\sim$~10~nm film on top of a 25~$\mu$m foil. Further studies to establish the effect of film thickness and other deposition parameters are underway, and will be published elsewhere.

One possible source of uncertainty in these measurements is  measurement of the pumping speed ($\dot{V}$), which relies on the accuracy of our ion gauge. The pumping speed was measured by observing the change in pressure with our ion gauge when a calibrated leak was opened to the system.  However, since the same pumping speed and experimental configuration was used for each measurement, we don't expect uncertainty in the pumping speed to affect the determination of the permeation mechanism, or the calculation of the PRF. 

\begin{table}
\begin{tabular}{|c|c|c|}
    \hline
     Sample & Prefactor (mol/(MPa$^{1/2}$*s*m))& Activation Energy (kJ/mol)\\\hline
     As-received Cu & 1.2x10$^{-4}$ & 75.8 \\\hline
     Annealed Cu & 4.5x10$^{-4}$ & 74.7 \\\hline
      - & Prefactor (mol/(MPa*s*m)) & - \\\hline
     ALD Film & 1.3x10$^{-4}$ & 77.4 \\\hline
\end{tabular}
    \caption{Results of the Arrhenius analysis of the permeation.}
    \label{tab:Permeation}
\end{table}

\begin{table}
\begin{tabular}{|c|c|}
    \hline
     Temperature (\degC) & Permeation Rate Reduction Factor \\
     & 100 Torr (unitless)\\\hline
     350 & 22.7 \\\hline
     325 &  21.2\\\hline
     300 &  30.9\\\hline
     275 &  34.3\\\hline
\end{tabular}
    \caption{Temperature dependence of the permeation rate reduction factor, calculated by the equation (Perm Rate)$_\textrm{annealed}$/(Perm Rate)$_\textrm{ALD+anneal}$. Because of the surface vs bulk limiting permeation modality, these numbers were calculated at fixed upstream pressure of 100 Torr.}
    \label{tab:PRF}
\end{table}

\section{Conclusions}
Atomic layer deposition and hydrogen permeation instruments have been developed for measurement of materials relevant to fusion energy. The process window for deposition of high quality ALD \ce{Al2O3} was established on silicon, and used to deposit films on Cu foil substrates. The \ce{Al2O3} films significantly affected the permeation of \ce{D2} through the Cu, decreasing the molar flux by a factor of more than 20, and fundamentally changing the permeation mechanism. Future work will explore film properties using the systems developed in this paper, and how those properties impact permeation.

\begin{acknowledgments}
This material is based upon work supported by the Department of Energy [National Nuclear Security Administration] University of Rochester “National Inertial Confinement Fusion Program” under Award Number(s) DE-NA0004144.

The co-authors would like to thank Greg Amos and Tim Clark and their team of machinists for support building many of the experimental setups used in this study.

The co-authors would like to thank Scott Householder and his team of safety officers for making sure that our home-built systems operate safely and are in compliance with all of the Laboratory for Laser Energetics' safety policies. 

Co-author Robinson would like to thank Dr. Luke Ceurvorst for useful conversations about characterizing the ALD system's raw data files, and Professor Wyatt Tenhaeff for access to ellipsometry.

The data that support the findings of this study [specific details here] are openly available in the [insert location here, such as arXiv.org, or University of Rochester Research Repository (URRR), an instance of Figshare], at [ link DOI here ]

\end{acknowledgments}

\pagebreak
\section{Bibliography}
\bibliography{Bibliography}
\pagebreak
 
\section{Figures}

\end{document}